\documentclass[superscriptaddress,twocolumn,showpacs,a4paper,
amssymb,amsmath,nobibnotes,aps,prd,
showkeys,
nofootinbib,notitlepage]{revtex4-1}
\usepackage{verbatim}
\usepackage[T1]{fontenc}
\usepackage[utf8]{inputenc}
\usepackage[american]{babel}
\usepackage{epsfig}
\usepackage{ulem}
\usepackage{graphicx}
\usepackage{booktabs}
\usepackage{multirow}
\usepackage{dcolumn}
\usepackage{amsmath}
\usepackage{mathtools}
\usepackage{amsfonts}
\usepackage{amssymb}
\usepackage{epstopdf}
\usepackage{bm}
\usepackage{siunitx}
\usepackage{braket}
\usepackage{enumitem}
\usepackage{soul}
\usepackage[usenames,dvipsnames]{color}
\usepackage[table]{xcolor}
\usepackage{color}
\usepackage{transparent}
\usepackage[caption=false,]{subfig}
\usepackage{pifont}
\definecolor{navyblue}{rgb}{0.0, 0.0, 0.5}
\definecolor{royalblue}{rgb}{0.25, 0.41, 0.88}
\definecolor{blue-violet}{rgb}{0.54, 0.17, 0.89}
\definecolor{darkviolet}{rgb}{0.58, 0.0, 0.83}

\usepackage{hyperref}
\hypersetup{
	colorlinks=true, % false: boxed links; true: colored links
	linkcolor=royalblue, % color of internal links (change box color with linkbordercolor)
	citecolor=magenta}
\usepackage{booktabs}
\usepackage{multirow}
\usepackage{dcolumn}
\usepackage{colortbl}
\newcommand\vertsp{\rule[-2mm]{1mm}{0mm} &}
\newcommand\horsp{\rule[-1.5mm]{0mm}{4.125mm}}
\newcommand\morehorsp{\rule[-2.25mm]{0mm}{6mm}}

\newcommand{\ie}{{\it i.e. }}

\begin{document}

\title{Hubble speed from first principles}

\author{Fabrizio Renzi}
\email{renzi@lorentz.leidenuniv.nl}
\affiliation{Lorentz Institute for Theoretical Physics, Leiden University, PO Box 9506, Leiden 2300 RA, The Netherlands}
\author{Alessandra Silvestri}
\affiliation{Lorentz Institute for Theoretical Physics, Leiden University, PO Box 9506, Leiden 2300 RA, The Netherlands}

\begin{abstract} 
	We introduce a novel way of measuring $H_0$ from a combination of independent geometrical datasets, with no need of calibration nor of  the choice of a cosmological model. We build on the {\it distance duality relation} which sets the ratio of the luminosity and angular diameter distance to a fixed scaling in redshift for any metric theory of gravity with standard photon propagation  and constitutes a founding block of any theory describing our Universe. Our method provides the unprecedented possibility of determining $H_0$ from first principles, unleashing the measurement of this fundamental constant from calibration and assumption of a cosmological model. We find $H_0=69.5 \pm 1.7$ km/s/Mpc at $68\%$ C.L. showing that the Hubble constant can be constrained at percent level with minimal assumptions.\end{abstract}

\maketitle

\section{Introduction}

 The Hubble parameter, $H_0$,  is arguably the most fundamental constant of our Universe. It provides its current expansion rate, an indication of its age and an overall scale for distances. After almost a century of increasingly precise determinations~\cite{Verde:2019ivm},  two leading experiments nowadays report values of $H_0$ in tension at more than $4\,\sigma$: $H_0 = 74.03 \pm 1.42$ km/s/Mpc at $68\%$ C.L.  from Type Ia supernovae (SNIa) calibrated using Cepheid variables stars within the SH0ES collaboration~\cite{Riess:2019cxk}\footnote{This result has recently been updated by the SH0ES collaboration with the new parallax measurements of nearby Cepheids from the GAIA satellite \cite{Riess:2021jrx}. The new constraint on the Hubble constant slightly improved in accuracy leading to $ H_0 = 73.30 \pm 1.04 $ km/s/Mpc . However the improved data have not yet been made public,  therefore  we continue to refer to the results presented in~\cite{Riess:2016jrr} in order to allow for a more informed comparison of our results. Comparing our findings with the latests SH0ES result would go in the direction of raising the statistical significance of  the discrepancy with our $H_0$ to around $ 0.5\sigma $} and $H_0=67.4 \pm 0.5$ km/s/Mpc $68\%$ C.L. inferred by Planck from the Cosmic Microwave Background (CMB), within the standard cosmological model $\Lambda$CDM~\cite{Aghanim:2018eyx}. Some yet unaccounted for systematics in CMB~\cite{DiValentino:2016hlg,Aghanim:2018eyx,Calabrese:2008rt,DiValentino:2019qzk,Handley:2019tkm} or SNIa~\cite{Mortsell:2021nzg,Mortsell:2021tcx,Efstathiou:2020wxn,Anderson:2021fsp,Freedman:2021ahq} data or a modification of the standard cosmological model (see e.g. for recent reviews~\cite{DiValentino:2021izs,Vagnozzi:2019ezj}) could reconcile these measurements, however no compelling resolution in either direction has been found yet. 
 
 A recent reanalysis of the Planck data \cite{Efstathiou:2019mdh} has shown that the results on cosmological parameters inferred from the Planck likelihood show no evidence of being affected by any significant systematic error. Also, results from recent CMB ground experiments have obtained  results consistent with Planck, assuming $\Lambda$CDM~\cite{ACT:2020gnv,ACT:2020frw,SPT-3G:2021eoc,SPT-3G:2021wgf} \footnote{It is worth noting that those experiments employ a prior information on the cosmic optical depth coming directly from Planck. Therefore they cannot be considered completely independent from the Planck measurements.}. The SH0ES team has also performed extensive analysis of possible systematic errors in the SNIa calibration procedure or in the Cepheids measurements, and their impact on the determination of $H_0$~\cite{Riess:2016jrr,Riess:2019cxk,Riess:2021jrx} excluding a full resolution of the tension with Planck via a systematic.
 The SH0ES collaboration uses the light curves of variable Cepheid stars to anchor SNIa and infer $H_0$.  While these methods are robust,  some unaccounted-for systematics could be causing the value of SH0ES to be in tension with Planck~\cite{Efstathiou:2020wxn}. Different methods of calibration, using either the surface brightness fluctuations of SNIa (SBF)  or the tip of the Red Giant  Branch (TRGB) have respectively measured a value of $H_0 = 70.50 \pm 5.75$ km/s/Mpc~\cite{Khetan:2020hmh} and $H_0 = 69.6 \pm 1.9$ km/s/Mpc~\cite{Freedman:2020dne}. 
 
The  expansion rate of the Universe sets the scaling of distances with time. In a homogeneous and isotropic Universe, described by the Friedmann-Lemaitre-Robertson-Walker (FLRW) metric, $ds^2 = -dt^2 + a^2(t)d\chi^2$ (where we set $c=1$), the comoving distance $\chi$ is related to the evolution of the scale factor $a(t)$ via $d\chi/dz = H^{-1}(z)$, where  we have introduced the redshift $z$ via the relation $a = (1+z)^{-1}$, and $H(z)$ is the Hubble parameter, i.e. the rate at which the Universe is expanding at a given time $H(t)\equiv d \ln a/dt$ expressed in terms of $z$, with $H(z=0)=H_0$ today. 

The evolution of the Hubble parameter depends on the cosmological model, and to highlight this we write: 
\begin{equation}\label{eq.Hubblelaw}
\chi(z) = \frac{1}{H_0}\int_{0}^{z}\frac{d\tilde{z}}{E(\tilde{z})}\,,
\end{equation}
where 
\begin{equation}\label{eq.dimensionless_H}
E^2(z)\equiv\frac{H^2(z)}{H^2_0}=\Omega_m (1+z)^3+\Omega_r (1+z)^4+\Omega_X X(z)\,.
\end{equation}
With $ \Omega_X $ we broadly indicate any contribution to $ H(z) $ not coming from matter or radiation e.g. $\Lambda$CDM with non-zero spatial curvature would give $\Omega_X\neq0$. The possible time dependence of any kind of dynamical dark energy or modification of gravity will be captured by $ X(z) $, with $X=1$ corresponding to the cosmological constant $\Lambda$~\cite{Wang:2018fng}.
%
%the cosmological constant, a dynamical dark energy or modifications of gravity.   The possible time dependence is captured by $X(z)$. In practice, this parameter encodes any contribution to $H(z)$ not coming from matter or radiation, e.g. $\Lambda$CDM with non-zero spatial curvature would give $X\neq1$. 

These equations show that reconstructing the evolution of cosmic distances in redshift, we can probe not only into the dynamics of the Universe, but also into its composition or, more broadly, its {\it cosmological model} $E(z)$. They also highlight how the inference of cosmological parameters from distances is inevitably characterized by a strong correlation between $H_0$ and the cosmological model, which is more severe at higher redshift. That is why the so-called early time measurements of $H_0$, e.g. CMB, are model dependent. Low redshift, late-time measurements such as SNIa, probe $H_0$ more directly, as can be seen from  the $z\rightarrow 0$ limit of Eq.~(\ref{eq.Hubblelaw}); yet they crucially depend on an overall calibration of nearby distances. Wanting to shed light on the Hubble tension, we seek a way to constrain $H_0$ independently of any assumption on the cosmological model and calibration.

\section{From distances to the Hubble parameter}

Cosmological observations measure the flux from sources of known intrinsic brightness, inferring the {\it luminosity distance}, $d_L$, or the angular scale of an object of known size, inferring the {\it angular diameter distance}, $d_A$.  They are both related to the comoving distance, albeit in a model-dependent way. On more general terms, in any metric theory of gravity, if photons propagate along null geodesics and obey  number conservation, it is possible to show that at any redshift
\begin{equation}\label{eq.eta_theory}
\eta(z) \equiv \frac{d_L(z)}{(1+z)^2d_A(z)} = 1\,,
\end{equation}
based solely on general geometrical arguments~\cite{Etherington}.
Eq.~(\ref{eq.eta_theory})  is known as the {\it distance duality relation} (DDR). We focus, quite generally, on theories for which the DDR holds, which means frameworks with a metric theory of gravity and standard photon propagation. Hence, for us $\eta$ is set to unity by the theory. On a side note, let us notice that experimental constraints have been placed on $\eta$,  that allow only tiny deviations from unity at the order of $10^{-2} - 10^{-1}$~\cite{Zhou:2020moc,Holanda:2020fmo,Xu:2020fxj,Holanda:2015zpz,Holanda:2016msr,Rana:2017sfr}. 

Setting $\eta=1$,  Eq.~(\ref{eq.eta_theory}) can then be rewritten as 
\begin{equation}\label{eq.H0_DDR}
H_0 =\frac{1}{(1+z)^2} \frac{H_0d_L(z)}{H(z)d_A(z)}H(z) \,.
\end{equation}
The term at the denominator, $H(z)d_A(z)$, can be obtained from a combination of line-of-sight and transverse BAO measurements, without the need of an external anchor for the sound horizon $r_s$ at drag epoch (as we will discuss in more detail in Section \ref{sec.data_methods}). Cosmic chronometers (CC) are standard clocks which, combining differential age estimates of systems with passively evolving star populations (e.g. globular clusters) with their corresponding spectroscopic redshift, determine $H(z)=-(1+z)^{-1}\Delta z/\Delta t$. As such, they  provide $H(z)$ data that is free from calibration and does not depend on the underlying cosmological model~\cite{Zhang:2014,Stern:2010,Moresco:2012,Moresco:2016mzx,Moresco:2015cya}. As we discuss in Appendix \ref{App.CCH}, they carry some dependence on the astrophysical modeling. Finally,  $H_0d_L$ can be obtained  from measurements of SNIa without the need of calibration. Combining BAO and CC first, and then folding in SNIa we obtain:
\begin{equation}\label{eq.eta_CC_SNIA_BAO_actual}
H_0=\frac{1}{(1+z)^2} \frac{\left[H_0d_L(z)\right]^{\rm SNIa}}{\left[d_A(z)\right]^{\rm BAO+CC}}\,.
\end{equation}

We get $\left[H_0d_L(z)\right]^{\rm SNIa}$ directly from the measurements of apparent magnitude through $  m_B  =5\log_{10}(H_0d_L) -5a_B$ (where we fix $a_B = 0.71273 \pm 0.00176$ as inferred in~\cite{Riess:2016jrr} directly from the same SNIa measurements that we are using, and independently of CMB and BAO). The values of $ m_B $ we used in this work are the standardized apparent peak magniturde of SNIa as listed in the Pantheon catalogue. This is an heavily processed quantity derived from analyzing the light curves of each SNIa in the catalogue. The Pantheon catalogue contains thousands of such measurements collected from different experiments and reanalyzed to smooth over the peculiar features of each survey (for more details about the SNIa standardization process we refer to \cite{Scolnic:2017caz} and references there in).
It is worth noting that using this value of $ a_B $ bears some minimal model dependence in the choice of the form of the expansion rate. Such assumptions have been reviewed in the latest SH0ES release~\cite{Riess:2021jrx} showing no sign of inconsistency with the assumptions made in \cite{Riess:2016jrr}.
Using the combination $H_0d_L$  we do not need  an estimate of the absolute magnitude $M_B$ and are thus free from calibration.
\begin{figure}
	\centering
	\includegraphics[width=0.49\textwidth,keepaspectratio]{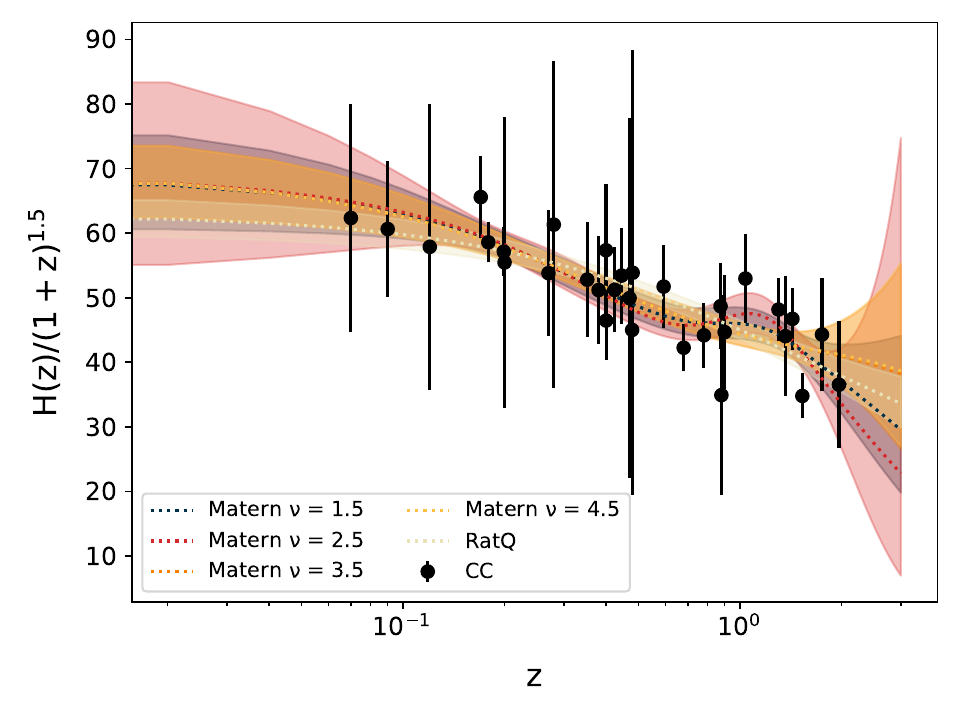}
	\includegraphics[width=0.49\textwidth,keepaspectratio]{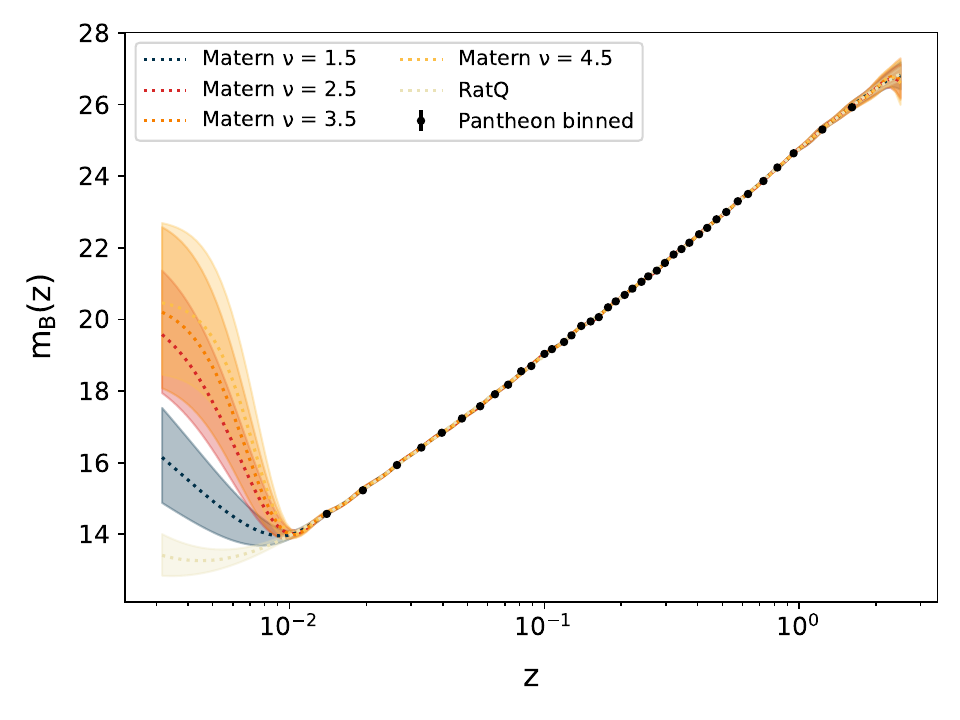}
	\caption{GP interpolations of the the CCH (upper panel) and SNIa (bottom panel) data with different kernels from the Matern family (\autoref{eq.Matern}) as well as with the RatQ kernel (\autoref{eq.RatQ}). The fits are shown against the data (black dots), which in the SNIa case are the binned Pantheon data. Different colors correspond to different kernels, as shown in the legend. For each kernel, the solid line represents the mean, and the three corresponding shaded regions represent the $1 \sigma$ level.}
	\label{fig.kernels_comparison_CCH}
\end{figure}

In our approach, assuming the validity of the DDR is equivalent to assuming that all probes are consistent tracers of cosmic expansion. If this was not the case, the $H_0$ inferred from Eq.~(\ref{eq.eta_CC_SNIA_BAO_actual}) would not be a constant, but it would rather show some, unphysical trend in redshift. The reformulation of the DDR in terms of Eq.~(\ref{eq.eta_CC_SNIA_BAO_actual}),  makes manifest the role of $H_0$ not only as an absolute distance scale, but also as a trigger of inconsistencies among distance probes. 
In particular an inconsistency in $H_0$ from two  experiments probing the same redshift, corresponds necessarily to an inconsistency in their distances. In fact the $~2\sigma$ discrepancy between the value of $H_0$ from TRGB and SH0ES can be linked to an inconsistency at $2\sigma$ in their measures of distances to common SNIa hosts~\cite{Efstathiou:2020wxn}. Other minimally model-dependent methods to estimate $H_0$ have been proposed in the literature in recent years. For example, methods that employ interpolation of the expansion history $H(z)$ and extrapolate towards $H_0$~\cite{Bernal:2016gxb,Heavens:2014rja,Verde:2016ccp,Pogosian:2020ded,Benndorf:2021lke} or employ the inverse distance ladder~\cite{Aubourg:2014yra,Cuesta:2014asa,Taubenberger:2019qna,Camarena:2019rmj}. Typically these methods carry still some dependence on the value of the drag scale $r_s$~\cite{Camarena:2019rmj,Pogosian:2020ded} and/or other cosmological parameters (e.g. curvature density~\cite{Aubourg:2014yra,Cuesta:2014asa,Taubenberger:2019qna,Bernal:2016gxb,Heavens:2014rja,Verde:2016ccp,Benndorf:2021lke}). With our approach,  we further remove such dependences and effectively only rely on the assumption of the validity of the DDR.

\section{Data sets and Methodology}\label{sec.data_methods}

We use three publicly available datasets:  the Pantheon dataset~\cite{Scolnic:2017caz}, a collection of around one thousand SNIa relative brightness measurements;  seven BAO data points from the last data release of the BOSS collaboration (see~\autoref{tab:BAOdata})  and a compilation of $30$ CC measurements of $H(z)$ in the interval $0<z<2$~\cite{Zhang:2014,Stern:2010,Moresco:2012,Moresco:2016mzx,Moresco:2015cya} (see \autoref{tab.CosmicChrono}).  

These datasets cover almost the same range of redshift ($0 < z \lesssim 2 $), therefore they measure the very same volume of Universe with three independent methods.   Yet, they are discrete and provide measurements of $H(z)d_A(z)$, $H_0d_L(z)$ and $H(z)$ at different redshifts. We choose to work with the seven redshift points corresponding to the BAO dataset, and apply a Gaussian process (GP) regression to fit to SNIa and CC in order to reconstruct them at the chosen redshifts. At each redshift, the GP regression estimates the expected value of the reconstructed function and its confidence interval, providing us with a continuous set of {\it probability distribution functions} (PDFs) at any given redshift. The outcome of this procedure is therefore a multidimensional probability distribution in the functional space of $H_0d_L(z)$ and $H(z)$, predicting the behaviour of these functions where data are not available. In~\autoref{fig.kernels_comparison_CCH} we show the outcome of the GP fits to the $H(z)$ data from CC and to the (binned) magnitude data of SNIa, with different kernels shown in different colors. One can clearly notice that all kernels give a very good representation of the measurements in the range where data exist, with negligible differences among the different kernels. However, outside the data range, the fit becomes unreliable, with, in some cases, severe differences among the different kernels, highlighting the biasing potentially introduced by extrapolations. In our analysis we rely on the GP fit strictly within the redshift range of data. In the latter, we choose the representation of the GP that gives the lowest $ \chi^2 $ (calculated as described in Appendix~\ref{App.GPMC}), i.e. the Rational Quadratic (RatQ) for the CCH measurements and the Matern kernel for the SNIa data with shape parameters $ \nu = 2.5 $. For completeness we report here the analytical form of the RatQ kernel \cite{GPbook} :
\begin{equation}\label{eq.RatQ}
	\kappa(z_i,z_j)_{\rm RatQ} = \sigma_M^2\left(1 + \frac{d(z_i,z_j)}{2\alpha\ell}\right)^{-\alpha}
\end{equation} 
and the general form of the Matern kernel:
\begin{equation}\label{eq.Matern}
	\kappa(z_i,z_j) = \sigma_M^2\frac{\left(\frac{\sqrt{2\nu}}{\ell}d(z_i,z_j)\right)^\nu}{\Gamma(\nu)2^{\nu - 1}} K_\nu\left(\frac{\sqrt{2\nu}}{\ell}d(z_i,z_j)\right)
\end{equation}
where $ K_\nu(\cdot) $ and $ \Gamma(\cdot) $ are the modified Bessel and Gamma functions while $ d(z_i,z_j) $ is the Euclidean distance between $ z_i $ and $ z_j $.
In both cases we fit the hyperparameters of the kernels $ \{\sigma_M,\ell,\alpha \} $ directly to the data minimizing a Gaussian likelihood (see  Appendix~\ref{App.GPMC} for more details).

\begin{figure}[t]
	\centering
	\includegraphics[width=0.49\textwidth,keepaspectratio]{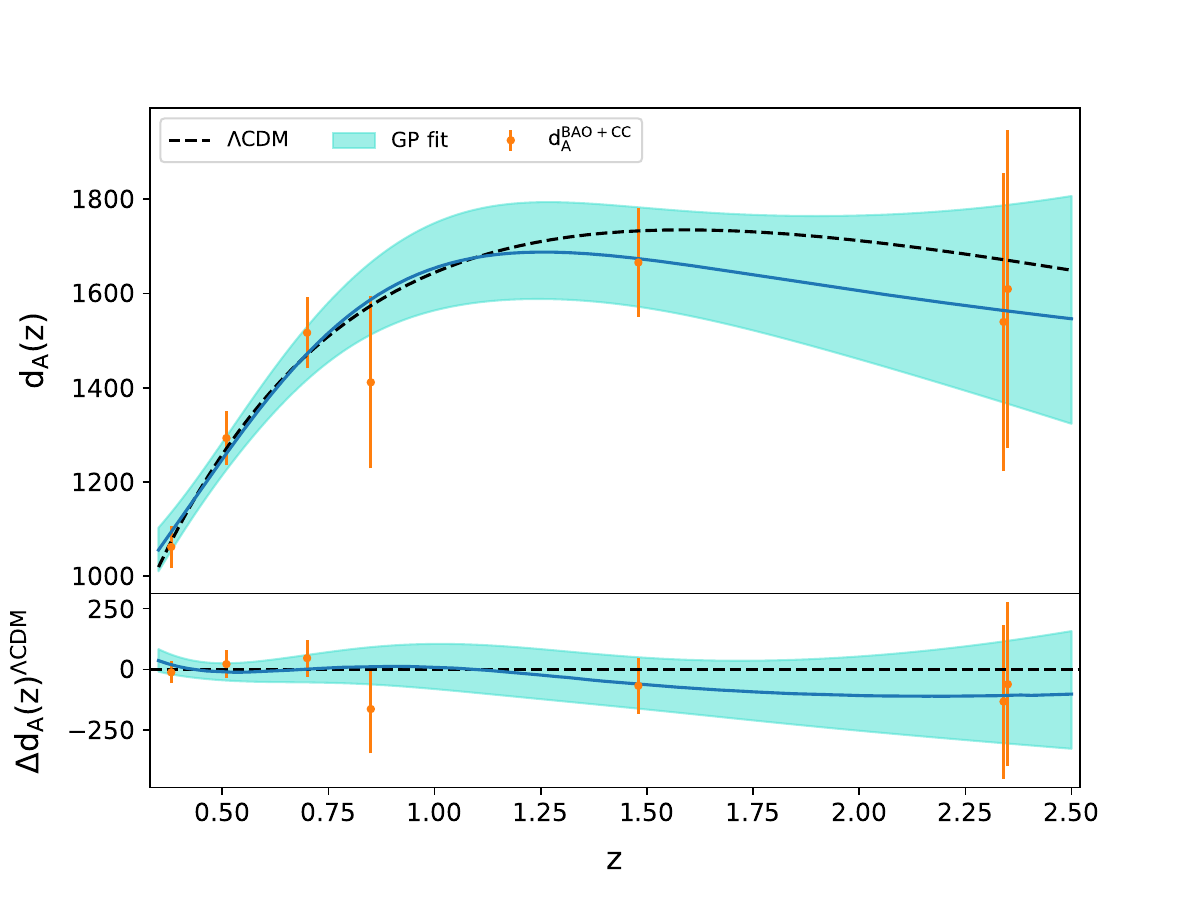}	\includegraphics[width=0.49\textwidth,keepaspectratio]{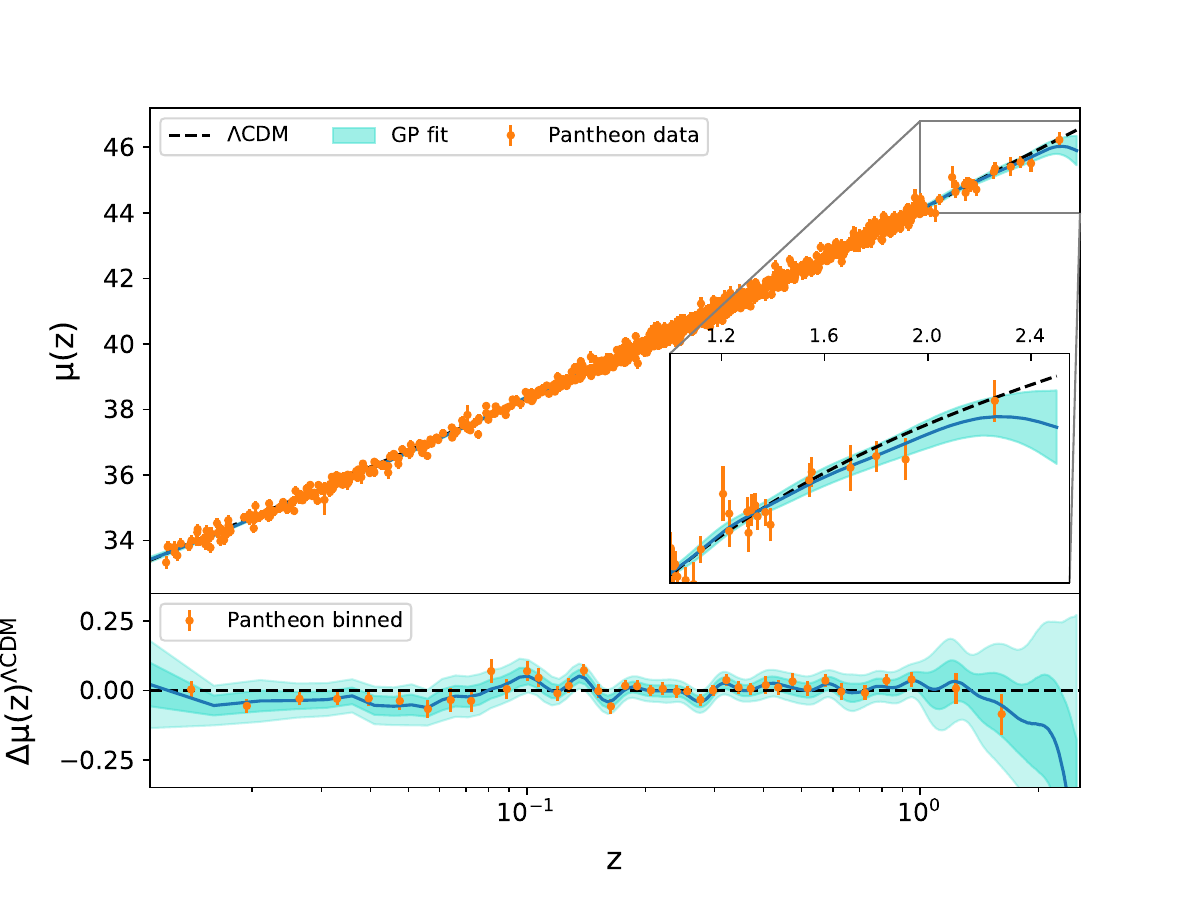}
	\caption{{\bf top:} Angular diameter distance reconstructed  from  BAO and CC data, and corresponding residual; the points with error bars correspond to the data reconstructed via the GPMC applied to the  BAO and CC measurements; the solid blue curve is the GP fit to the data points. {\bf bottom:} GP fit to the luminosity distance from  SNIa (Pantheon) data. In both panels, the dashed black curve shows the $\Lambda$CDM model predictions corresponding to our best-fit $H_0$ and $\Omega_m = 0.3$, with respect to which the residuals are calculated. }
	\label{fig.DA_DL} 
\end{figure}

The resulting GP fit to the luminosity distance of SNIa is shown in the bottom panel of~\autoref{fig.DA_DL}, against the Pantheon data and the $\Lambda$CDM prediction.

From these fits, we extract samples of the PDF of $H_0d_L(z)$ and $H(z)$ at the BAO redshifts.  In other words, we extract a number of realizations compatible with the statistics dictated by the distribution function of the GP fit.  Specifically, we draw  $10000$ realizations of  $H_0d_L(z)$ and $10000$ realizations of the expansion rate $H(z)$.  
As for the BAO data, we do not perform any GP fit, and we use their seven redshift points as the reference ones. We combine the BAO data in Tab.\ref{tab:BAOdata}, i.e. $d_M/r_s$ and $d_H/r_s$, assuming they have gaussian PDFs (based on the symmetry of their $68\%$ confidence intervals), to obtain PDFs of $H(z)d_A(z)$ at the seven redshifts. 

We proceed combining the realizations of CC $H(z)$ (from the GP fit) with the unanchored BAO data $H(z)d_A(z)$ to obtain $10000$ samples of the angular distance, $\left[d_A(z)\right]^{\rm BAO+CC}$ at the seven BAO redshifts. We show the corresponding estimates in the upper panel of~\autoref{fig.DA_DL} (orange dots), along with the  $\Lambda$CDM prediction for $d_A(z)$ and a simple GP fit to $\left[d_A(z)\right]^{\rm BAO+CC}$. We do not use this latter fit in the subsequent steps, still it provides a useful check for any deviations with respect to the standard cosmological model in our reconstruction of the BAO distance. 

While this procedure may appear similar to the standard way of calibrating BAO with external information on the sound horizon $r_s$,  the CC data directly provides estimates of the value of $H(z)$ at several redshifts, without any dependence on the cosmological model. 

It is certainly true that combining the $d_M$ and $d_H$ data, one effectively reduces the constraining power of BAO, however one gains independence from calibrations of the sound horizon or assumptions on the cosmological model. Any other way of using the BAO data to infer $H_0$ would include one, or more, of the latter. For instance, one could calibrate the BAO directly using the CC measurements as done e.g. in~\cite{Heavens:2014rja,Verde:2016ccp}, but this requires to make an assumption about the form of the expansion history and to include the curvature density in the analysis. Another possibility would be to  constrain  the combination $ r_sH_0 $ (that can be fixed without including high redshift information and the curvature density~\cite{Pogosian:2020ded}), however this would not provide a measure of the Hubble constant independent of the value of the sound horizon. Finally, as proposed in \cite{Schoneberg:2019wmt,DES:2017txv}  it is possible to combine BAO data and Big Bang Nucleosynthesis (BBN) to estimate the sound horizon $r_s$. However BBN codes depend on the knowledge of the value of the neutron lifetime, which is used to calculate the abundance of primordial elements. Currently there is a $ 4\sigma $ discrepancy in the measurements of the neutron lifetime between bottle-only and beam-only experiments (see e.g. ~\cite{Salvati:2015wxa,ParticleDataGroup:2020ssz})$  $. A change in the neutron lifetime can significantly bias the constraints from BAO data (particularly the value of $ N_{\rm eff} $ \cite{Capparelli:2017tyx}) and lead to a biased $ H_0 $ inference.     

\begin{figure}[tbp!]
	\centering
	\includegraphics[width=0.49\textwidth,keepaspectratio]{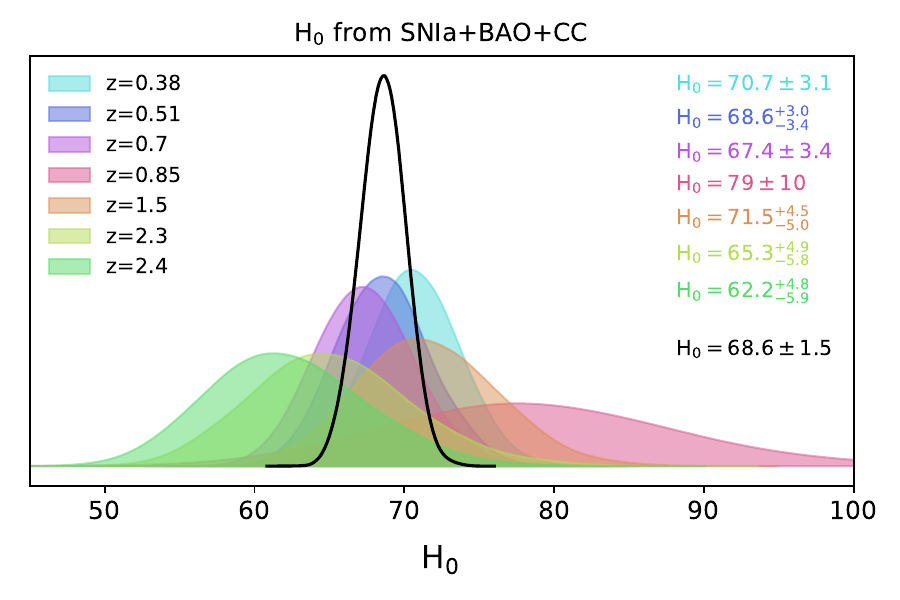}	\includegraphics[width=0.465\textwidth,keepaspectratio]{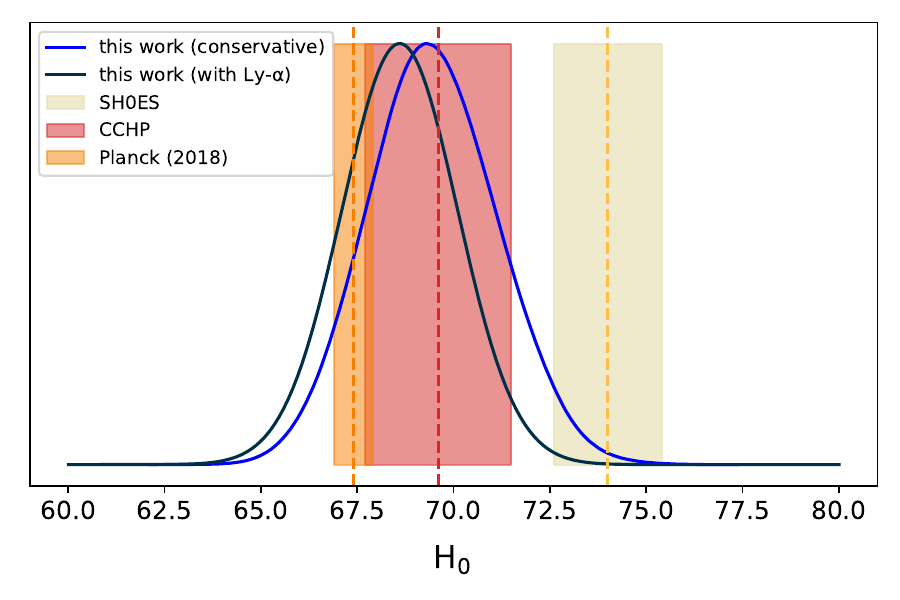}
	\caption{\textbf{top:} Posterior distribution functions (PDFs) for the Hubble parameter obtained from Eq.~(\ref{eq.eta_CC_SNIA_BAO_actual}), at the seven redshift points of BAO data (shaded curves). The solid black line represents the combined PDF (full result). \textbf{bottom:} PDF for $H_0$ corresponding to our full result (solid black curve), {\it conservative} result (solid blue line), along with recent constraints from SH0ES~\cite{Riess:2019cxk}, TRGB~\cite{Freedman:2020dne} and Planck~\cite{Aghanim:2018eyx}; shaded regions represent the 1$\sigma$ interval.}
	\label{fig.h0_post}
\end{figure}

Finally we combine the  $10000$ realizations of the uncalibrated $\left[H_0d_L(z)\right]^{\rm SNIa}$ with  the realizations of $\left[d_A(z)\right]^{\rm BAO+CC}$ by mean of Eq.~(\ref{eq.eta_CC_SNIA_BAO_actual}) to get an estimate of the PDF of $H_0$ at each of the BAO redshifts, ~\autoref{fig.h0_post}.

With the above procedure, the uncertainty on  $[H_0d_L]^{\rm SNIa}(z)$ and $[d_A]^{\rm BAO+CC}(z)$, as well as their mutual correlation are intrinsically included in the respective PDFs. We  base our error propagation on the PDF samples (see also ~\cite{Renzi:2020bvl}) to take advantage of this feature. For every sample of the PDFs of $[H_0d_L]^{\rm SNIa}(z)$ and $[d_A]^{\rm BAO+CC}(z)$, we calculate the corresponding sample of the posterior of $H_0$ through Eq.\eqref{eq.H0_DDR}. We then use these samples to reconstruct the multi-variate distribution of the 7 $H_0$ measurements. From the latter,  we extract the {\it marginal} distribution of the individual $H_0$ thus leaving us with 7 uncorrelated PDFs. 
This is analogous to the procedure used in MCMC to derive constraints on the likelihood parameters and allows to translate the correlations and the uncertainties of the PDFs of the distances directly into the PDFs of $H_0$. This guarantees that the correlations between individual $H_0$ and their errors are accounted for, and it further allows us not to make any specific assumption about the form of the PDFs of $H_0$ (e.g. approximating them to Gaussian PDFs).

Finally, we multiply together all the marginalized PDFs of $H_0$ into a single PDF from which one can easily extrapolate mean and variance by employing an inverse transform sampling. The individual $H_0$ PDFs and the final results obtained combining them are reported in Fig.\eqref{fig.h0_post}

The overall method is a combination of a MCMC-like parameter estimation and a Gaussian process reconstruction and we refer to it as Gaussian Process Monte Carlo (GPMC). It uses the least possible number of theoretical assumptions to derive cosmological information and provide results that are conditioned only by the GPMC reconstruction. We carefully removed any external calibration of our dataset as well as any assumption on the cosmological model. Further, we do not perform any extrapolation of the GP fit to $z\rightarrow 0$ in order to obtain $H_0$, but rather combine data in the range where they exist and estimate $H_0$ from the consistency of distances.  An extrapolation to $z\rightarrow 0$,  would be extremely sensitive to the choice of kernel, and would not necessarily  be a fair representation of the expansion rate  in the redshift range where data does not exist~\cite{Efstathiou:2021ocp,Camarena:2021jlr}. 

Because of these features, our methodology is different from approaches that use a parametrization of the cosmological model (e.g.~\cite{Aghanim:2018eyx,DiValentino:2016hlg,Dutta:2018vmq,Park:2018tgj,Chen:2016uno,DiValentino:2017iww,DiValentino:2020naf,Vagnozzi:2020dfn}), employ data that need to be calibrated with external information (e.g. anchored SNe~\cite{Riess:2016jrr,Riess:2019cxk} and  BAO~\cite{Alam:2016hwk,Alam:2020sor,Agathe:2019vsu,Blomqvist:2019rah}). It also differs from previous papers that have used Gaussian process techniques, for instance applied to the CC data in combination with other probes, training the GP regressor directly on the expansion history data, $H(z)$, and then extrapolating the fit to redshift zero to obtain and estimate of $H_0$~\cite{Gomez-Valent:2018hwc,Bonilla:2020wbn,Yu:2017iju}. Strong lensing time delay measurements also offer a way to measure $H_0$, see e.g.~\cite{Wong:2019kwg,Collet:2019abc}, however the uncertainty on the final result is significantly enlarged (up to $10\%$) once one takes into account the uncertainties in the choice of the lens mass model~\cite{Birrer:2020tax}. 

\section{Results}
Combining all the seven reconstructed $H_0$, we derive a $2\%$ constraint on the value of the Hubble speed, $H_0 = 68.5 \pm 1.5 $ km/s/Mpc. Our result is competitive with current measurements from~\cite{Aghanim:2018eyx,Riess:2016jrr,Riess:2019cxk,Birrer:2020tax,Pesce:2020xfe}, and robust to a revision of the systematics in CC measurements~\cite{Moresco:2020fbm} as we discuss in Appendix \ref{App.CCH}. 
Our value for $H_0$ is in $2.5\sigma$ discrepancy with the value of SH0ES~\cite{Riess:2019cxk} and around $1\sigma$ away from the Planck result. In the following we will elaborate more on both these aspects, as they crucially unveil important insights on the calibration of SNIa and on possible new physics. 

From our value of $H_0$, we infer the SNIa absolute magnitude $ M_B = -19.355 \pm 0.054$,  in $2.5 \sigma$ tension with the calibrated magnitude of SH0ES \ie $M_B = -19.22 \pm 0.04$. This discrepancy can be traced back to a constant systematic offset in the SNIa calibration of $|\delta M| = 0.138 \pm 0.067$ consistent with the results of~\cite{Efstathiou:2020wxn,Mortsell:2021nzg,Camarena:2019rmj}.
Our result is in perfect agreement with the $H_0$ measurement from TRGB~\cite{Freedman:2019jwv,Freedman:2020dne} and suggests that the discrepancy with SH0ES may just be related to an unaccounted-for systematics in the SNIa calibration via Cepheids.

In obtaining our result for $H_0$,  we used all the BAO data, even though the Lyman-$\alpha$ points fall outside the redshift range of CC and SNIa  and are therefore more prone to  kernel-dependent results.  By looking at the means of the distributions in~\autoref{fig.h0_post}, it is clear that these two points give a very low mean value for $H_0$, and this effectively lowers the final, combined value. Their contribution to the total error is instead negligible. We checked that the drop in the value of $H_0$ is not a feature introduced by the kernel choice. Removing those two points,  we obtain a slightly higher value of the Hubble constant with a similar precision \ie $H_0 = 69.5 \pm 1.7$ km/s/Mpc. We will refer to this constraint as {\it conservative} and hereafter we will quote it as our main result. 

 In combining our measurements via the DDR, we implicitly assume that the different data sets trace the underlying expansion history {\it equally well}. An offset in their ability to catch features in  $H(z)$ would show up as an apparent redshift-dependence of $H_0$, in our case as a discrepancy between full and conservative result. CC are direct measurements of the expansion history, while SNIa are sensitive to the integral of $H(z)$ over the line of sight. We do expect therefore a slight lag in their tracing of $H(z)$, which indeed can be noticed comparing  our reconstructions of  the luminosity distance from SNIa to that of the angular diameter distance  from BAO+CC, see~\autoref{fig.DA_DL}.  One clearly notices that the BAO+CC reconstruction starts showing a mild departure from $\Lambda$CDM at around $z \sim 1$, while the SNIa reconstruction shows hints of deviations only around $z\sim1.5$. In our analysis this lag shows up as a drop in the binned values of $H_0$ for higher redshifts and we correspondingly find it as a mild feature in our reconstruction of $X(z)$. This falls well within the 1$\sigma$ confidence interval of the GP reconstruction, therefore it does not constitute any evidence of a deviation, rather just a trend in line with the findings of~\cite{Zhao:2017cud,Wang:2018fng,Risaliti:2018reu,Raveri:2019mxg}.  Recently some analyses of the Pantheon dataset  highlighted a possible transition of the SNIa magnitude at $z \lesssim 0.2$~\cite{Dainotti:2021pqg,Kazantzidis:2020xta,DiValentino:2020kha}. These trends are noticed in the low redshift tail, and their interpretation is possibly very different from the one of the higher redshift trend that we find. Finally, it is tantalizing to notice that our trend is of the same level of the shift between the late-time TRGB measurement of $H_0$ and Planck's one, and in fact it would bring the two values in even better agreement.  Were upcoming measurements to corroborate a discrepancy between CMB and TRGB inferences of $H_0$, a physical trend in $H(z)$,  would resolve it.
 
\section{Discussion and Conclusions}
In conclusion, we have introduced a new approach to the Hubble parameter which relies only on few foundational assumptions at the basis of the distance duality relation and, importantly,  uses a combination of geometrical data that do not need calibration nor the assumption of a cosmological model. We use GP regression techniques to fit the data {\it uniquely} in the redshift range where they exist, avoiding possibly biasing extrapolation of $H(z)$ to $z=0$. Our result of $H_0=69.5\pm1.7$ km/s/Mpc shows the possibility of measuring $H_0$ at the percentage level with the least possible assumptions. 

Whereas current data do not allow for a complete resolution of the Hubble tension, our method clearly shows that an adjustment in the calibration of the SNIa, i.e. setting $M_B =-19.355 \pm 0.054$, $~2.5\sigma$ lower than SH0ES calibration, brings local measurements in agreement. We also show that a mild dynamical feature in $X(z)$ at intermediate redshift would further lower the latter, bringing them in even better agreement with the early time measurement of Planck.

The innovative data analysis technique that we presented, built on a combination of Gaussian Process and Monte Carlo methods (GPMC),  is very promising in terms of leading to competitive constraints that rely on the least possible theoretical assumptions. In our application of it to $H_0$, it uniquely stands out for not relying on the assumption of a cosmological model nor on the calibration of data, providing a novel measurement of $H_0$ from first principles. This is crucial if one wants to shed light on the Hubble tension avoiding dynamical determinations of it.  Building on the DDR our work focuses on the role of $H_0$ as a fundamental constant that univocally sets distances in the nearby Universe. We do not make any assumptions about the cosmological model to fit cosmic distances, but rather we build our estimate on distance ratios. 

While we have focused on the Hubble parameter, our GPMC method can have broad applications in cosmology. It offers a novel and powerful alternative to the popular Monte Carlo Markov Chain (MCMC) technique, which has been pivotal for the estimation of cosmological parameters. With a key improvement: the GPMC method does not require the choice of any cosmological model, thus freeing us from theoretical biases which can be particularly limiting when tackling tensions among data sets. As such, it allows a more direct inference of the  underlying physics from data. An immediate example  is the possibility of  estimating  the sound horizon, $r_s$, independently on the cosmological model. This can be achieved combining SNIa distance measurements, marginalized over their absolute magnitude, with  transverse BAO measurements,  \ie $D_A(z)/r_s$ (see~\cite{Bernal:2016gxb,Heavens:2014rja,Verde:2016ccp,Pogosian:2020ded} for previous works in this direction).

\begin{acknowledgments}
We thank  Guadalupe Ca$\rm  \tilde{n} $as Herrera, Alice Garoffolo, Archisman Ghosh, ,William Giare', Wendy Freedman, Matteo Martinelli, Subodh Patil and Adam Riess for helpful discussions.  We acknowledge support from the NWO and the Dutch Ministry of Education, Culture and Science (OCW) (through NWO VIDI Grant No.2019/ENW/00678104 and from the D-ITP consortium).
\end{acknowledgments}

\appendix
\section{The GPMC method}\label{App.GPMC}
A Gaussian Process (GP) is a fully Bayesian approach for data smoothing which can reconstruct a function directly from data without assuming a specific parameterization for the function itself \cite{GPbook}. 
At each redshift $z$, the GP is a multivariate Gaussian with zero mean which implies one has only to select a concrete functional shape for the covariance function, called \emph{kernel}, of the GP.
Kernels are parametrized by a set of hyperparameters that must be fitted to the data. We do so relying on the maximization of the following $\chi^2$ likelihood:
\begin{equation}
-2 \ln \mathcal{L} = (\hat{Y} - Y)^{T}\Sigma^{-1}(\hat{Y} - Y)
\end{equation}
which contains the data covariance matrix $\Sigma$  for the data $\hat{Y}$ corresponding to a given observable. Here $Y$ are the GP reconstruction for the observable.  This approach allows us to avoid kernel-dependent results, typical of the marginal likelihood method (see e.g.~\cite{Pandey:2019yic,Gomez-Valent:2018hwc}).  For the Pantheon data we proceed by using the full datasets (1048 SNIa) for the GP reconstruction while we compute the likelihood on the binned dataset (30 binned SNIa). For the Cosmic Chronometers instead we use the collection of \autoref{tab.CosmicChrono} for the GP reconstruction and we compute the likelihood on a set of six measurements of $E(z)$ obtained combining the Pantheon dataset and the high redshift SNIa of the Multi-Cycle Treasury (MCT) program~\cite{Riess:2017lxs}.

Our methodology leads to results that are basically insensitive to the choice of the GP kernel in the range \ie $0.05 \lesssim z < 2$ where the data are abundant with a shift between the reconstructed distance that is much smaller than the 1$\sigma$ uncertainty of the GP reconstruction. The reconstruction at $z > 2$ may depend significantly on the kernel, however we find that this concerns only the error of the reconstruction and that the results obtained with different kernels are highly consistent with one another. 
We select therefore  the GP reconstruction associated to the kernel giving the lowest $ \chi^2 $. These are respectively a Rational Quadratic kernel for the CC data and a Matern kernel~\cite{GPbook} with parameter $\nu= 5/2$ for the Pantheon dataset (see also Eq.\eqref{eq.RatQ} and Eq.\eqref{eq.Matern}).

Once the GP reconstruction is complete, we can extract samples of the probability distribution function (PDF) of $Y_i = Y(z_i)$ at each redshift directly from the GP fit. This allows us to compute the values of $H_0d_L(z)$ (from SNe) and $H(z)$ (from the CC) at the BAO redshifts. Specifically, we draw $10000$ realizations of the expansion rate $H(z)$ from the GP fit to the CC data. We then combine these realizations with the unanchored BAO data $H(z)d_A(z)$ to infer the angular diameter distance, $d_A(z)$. Notice that  the CC data directly provides $H(z)$ at several redshifts, without any dependence on the cosmological model, so this is different from the standard anchoring. We draw $ 10000 $ realizations of the unanchored SNIa distance $H_0d_L(z)$ from the GP fit to the SNIa data. Finally, we combine the two reconstruction by mean of Eq.~\eqref{eq.eta_CC_SNIA_BAO_actual} in the main text,  to get an estimate of the PDF of $H_0$ at each of the BAO redshifts.

As a final step, we multiply together all the PDFs of $H_0$ into a single PDF from which one can easily extrapolate mean and variance by employing an inverse transform sampling. In this way, we take into account  the correlations between individual $H_0$ and their errors, without making any specific assumption about the form of the PDFs (e.g. approximating them to Gaussian PDFs).

\begin{table}[thbp!]
	\begin{tabular}{ccc ccc }
		\toprule 
		\horsp
		$z$ \vertsp $H(z)$  \vertsp Refs.  \vertsp $z$ \vertsp $H(z)$  \vertsp Refs.\\
		\hline
		\hline
		\morehorsp
		$0.07$ \vertsp $69.0 \pm 19.6$ \vertsp \cite{Zhang:2014} \vertsp $0.4783$ \vertsp $80.9 \pm 9.0$ \vertsp \cite{Moresco:2016mzx}\\ 
		$0.09$ \vertsp $69.0 \pm 12.0$ \vertsp \cite{Stern:2010} \vertsp $0.48$ \vertsp $97.0 \pm 62.0$ \vertsp \cite{Stern:2010}\\
		$0.12$ \vertsp $68.6 \pm 26.2$ \vertsp \cite{Zhang:2014} \vertsp $0.593$ \vertsp $104.0 \pm 13.0$ \vertsp \cite{Moresco:2012}\\
		$0.17$ \vertsp $83.0 \pm 8.0$ \vertsp \cite{Stern:2010} \vertsp $0.68$ \vertsp $92.0 \pm 8.0$ \vertsp \cite{Moresco:2012}\\
		$0.179$ \vertsp $75.0 \pm 4.0$ \vertsp \cite{Moresco:2012} \vertsp $0.781$ \vertsp $105.0 \pm 12.0$ \vertsp \cite{Moresco:2012}\\
		$0.199$ \vertsp $75.0 \pm 5.0$ \vertsp \cite{Moresco:2012} \vertsp $0.875$ \vertsp $125.0 \pm 17.0$ \vertsp \cite{Moresco:2012}\\
		$0.2$ \vertsp $72.9 \pm 29.6$ \vertsp \cite{Zhang:2014} \vertsp $0.88$ \vertsp $90.0 \pm 40.0$ \vertsp \cite{Stern:2010}\\
		$0.27$ \vertsp $77.0 \pm 14.0$ \vertsp \cite{Stern:2010} \vertsp 	$0.9$ \vertsp $117.0 \pm 23.0$ \vertsp \cite{Stern:2010}\\
		$0.28$ \vertsp $88.8 \pm 36.6$ \vertsp \cite{Zhang:2014} \vertsp $1.037$ \vertsp $154.0 \pm 20.0$ \vertsp \cite{Moresco:2012}\\
		$0.352$ \vertsp $83.0 \pm 14.0$ \vertsp \cite{Moresco:2012} \vertsp $1.3$ \vertsp $168.0 \pm 17.0$ \vertsp \cite{Stern:2010}\\
		$0.3802$ \vertsp $83.0 \pm 13.5$ \vertsp \cite{Moresco:2016mzx} \vertsp $1.363$ \vertsp $160.0 \pm 33.6$ \vertsp \cite{Moresco:2015cya}\\
		$0.4$ \vertsp $95.0 \pm 17.0$ \vertsp \cite{Stern:2010} \vertsp $1.43$ \vertsp $177.0 \pm 18.0$ \vertsp \cite{Stern:2010}\\
		$0.4004$ \vertsp $77.0 \pm 10.2$ \vertsp \cite{Moresco:2016mzx} \vertsp $1.53$ \vertsp $140.0 \pm 14.0$ \vertsp \cite{Stern:2010}\\
		$0.4247$ \vertsp $87.1 \pm 11.2$ \vertsp \cite{Moresco:2016mzx} \vertsp $1.75$ \vertsp $202.0 \pm 40.0$ \vertsp \cite{Stern:2010}\\
		$0.44497$ \vertsp $92.8 \pm 12.9$ \vertsp \cite{Moresco:2016mzx} \vertsp $1.965$ \vertsp $186.5 \pm 50.4$ \vertsp \cite{Moresco:2015cya}\\
		$0.47$ \vertsp $89.0 \pm 49.6$ \vertsp \cite{Ratsimbazafy:2017vga} \vertsp \\
		\bottomrule
	\end{tabular}
	\caption{The collection of $H(z)$ measurements from cosmic chronometers.}
	\label{tab.CosmicChrono}
\end{table}	

\begin{table*}
	\centering
	\begin{tabular}{ccccc}
		\toprule
		Type \vertsp $z$ \vertsp $d_M/r_s$ \vertsp $d_H/r_s$  \vertsp Reference  \\
		\hline\hline
		\morehorsp
		BOSS galaxy--galaxy \vertsp $ 0.38 $ \vertsp $ 10.27\pm 0.15 $ \vertsp $ 24.89\pm 0.58 $ \vertsp \cite{Alam:2016hwk} \\
		eBOSS galaxy--galaxy \vertsp $ 0.51 $ \vertsp $ 13.38\pm0.18 $ \vertsp $ 22.43 \pm 0.48 $ \vertsp \cite{Alam:2016hwk}\\
		\vertsp $ 0.70$ \vertsp $ 17.65 \pm 0.30 $ \vertsp $ 19.78\pm 0.46 $ \vertsp \cite{Bautista:2020ahg,Gil-Marin:2020bct}\\
		\vertsp $ 0.85 $ \vertsp $ 19.50\pm 1.00 $ \vertsp $ 19.60\pm 2.10 $ \vertsp \cite{Tamone:2020qrl,deMattia:2020fkb}\\
		\vertsp $1.48$ \vertsp $ 30.21\pm0.79 $ \vertsp $ 13.23\pm 0.47 $ \vertsp\cite{Neveux:2020voa,Hou:2020rse}\\
		eBOSS Ly-$\alpha$--Ly-$\alpha$ \vertsp $ 2.34 $ \vertsp $ 37.41 \pm 1.86 $ \vertsp $ 8.86\pm0.29 $ \vertsp\cite{Agathe:2019vsu} \\
		eBOSS Ly-$\alpha$--quasar \vertsp $ 2.35 $ \vertsp $ 36.30\pm1.80 $ \vertsp $ 8.20\pm0.36 $ \vertsp \cite{Blomqvist:2019rah} \\
		\bottomrule
	\end{tabular}
	\caption{The BAO data used in this work given in terms of $d_M = (1+z)d_A$ and $ d_H = c\,H(z)^{-1}$.}
	\label{tab:BAOdata}
\end{table*}

\section{Systematics of the CC measurements}\label{App.CCH}
In our approach,  SNIa and BAO data effectively constrain the evolution with redshift $E(z)$,   while CC set an absolute scale by  providing the value of $H(z)$ at each redshift. It is therefore important to asses thoroughly the systematics in the CC measurements that could affect our $H_0$ inference. 
CC measurement build on the following relation between $H(z)$ and the differential time-redshift relation: $H(z) = \frac{-1	}{1+z}\frac{\Delta z}{\Delta t}.$
While it is straightforward to measure $\Delta z$ through spectroscopic observations, measuring $\Delta t$ is a more challenging quest, which requires standard clocks, typically provided by differential ages of passively evolving stellar population. 

The possible sources of error in CC data come from uncertainties in modeling this stellar population~\cite{Moresco:2012by,Moresco:2015cya,Moresco:2016mzx,Moresco:2018xdr,Moresco:2020fbm}, and can be summarized as follows : the CC measurements require to select an unbiased tracer of the evolution of the differential age of the Universe with redshift and it is important to asses the impact of a subdominant young stellar population;  passive-evolving galaxies have been found to be an optimal tracer of the cosmic differential age yet they cannot be exactly described as composed by a single stellar population and one needs to asses the impact of assuming more realistic star formation history (SFH); the metallicity of the stellar population is often used as a prior in the calibration of the relative age of a stellar population, thus the uncertainties in modeling the stellar population metallicity must  be included in the total error budget; finally,  the CC measurements employ stellar population synthesis (SPS) model to calibrate the relative stellar age; the uncertainties coming from different modeling of the SPS are also a possible source of systematics. 

The CC data that we use for our main analysis (see~\autoref{tab.CosmicChrono}), already include the uncertainties associated with the SFH and the stellar metallicity~\cite{Moresco:2012by,Moresco:2015cya,Moresco:2016mzx}. The contribution of residual stellar population was shown to be negligible for this data set~\cite{Moresco:2018xdr}. Here,  we further consider the inclusion of SPS model uncertainties following the results of~\cite{Moresco:2020fbm}, where the latter were shown to contribute an additional $ \lesssim 16\% $ uncertainty in the CC measurements of $H(z)$, in the worst case scenario. To include this error we first reconstruct the evolution of the SPS error with redshift through a GP fit on the values reported in the third column of Tab.3 in~\cite{Moresco:2020fbm}. We then sum in quadrature the reconstructed error to that reported in~\autoref{tab.CosmicChrono} and we repeat our procedure to derive an estimate of $H_0$. We find that the effect of adding the SPS uncertainty is rather small, increasing the error of our fit by only $\sim 0.5 $ km/s/Mpc (consistently with~\cite{Vagnozzi:2020dfn}) and leading to $H_0 = 69.7 \pm 2.1$  km/s/Mpc in the conservative case and $ 68.9 \pm 2.1 $ km/s/Mpc when also Ly-$\alpha $ BAO are included in the fit. Our main conclusions remain therefore the same, with a $2\sigma$ shift with SH0ES that can be explained by a corresponding offset in the absolute magnitude, and a $\sim 1\sigma$ shift with Planck which may relate to a mild dynamical feature in $X(z)$.

%\section{A note on the conversion of $ M_B $ into $ H_0 $}
%\fabc{i think here we may need a new section to discuss the assumptions we made for aB in the main text.}

\bibliographystyle{aipnum4-1}
\bibliography{bib.bib}

\end{document}